\documentclass[pdflatex, sn-mathphys-num]{sn-jnl}

\usepackage{graphicx}
\usepackage{lipsum}  
\usepackage{multirow}
\usepackage{amsmath,amssymb,amsfonts}
\usepackage{amsthm}
\usepackage[title]{appendix}
\usepackage{xcolor}
\usepackage{textcomp}
\usepackage{manyfoot}
\usepackage{booktabs}
\usepackage{algorithm}
\usepackage{algorithmicx}
\usepackage{algpseudocode}
\usepackage{listings}
\usepackage{lmodern}
\usepackage[english]{babel}
\usepackage[T1]{fontenc}
\usepackage{float}
\usepackage{array}
\usepackage[utf8]{inputenc}
\usepackage{natbib}
\usepackage{hyperref}  

\begin{document}

\title[Article Title]{High-resolution ensemble retrieval of cloud properties for all-day based on geostationary satellite}




\author[1,2,3]{\fnm{Haixia} \sur{Xiao}}\email{haixiaxiao3@gmail.com}
\author*[1]{\fnm{Feng} \sur{Zhang}}\email{fengzhang@fudan.edu.cn}
\author[4]{\fnm{Lingxiao} \sur{Wang}}\email{lingxiao.wang@riken.jp}
\author[5]{\fnm{Baoxiang} \sur{Pan}}\email{panbaoxiang@lasg.iap.ac.cn}
\author[6]{\fnm{Yannian} \sur{Zhu}}\email{yannian.zhu@nju.edu.cn}
\author[6]{\fnm{Minghuai} \sur{Wang}}\email{Minghuai.Wang@nju.edu.cn}
\author[7]{\fnm{Wenwen} \sur{Li}}\email{liwenwen0213@foxmail.com}
\author[1]{\fnm{Bin} \sur{Guo}}\email{21113020009@m.fudan.edu.cn}
\author[8]{\fnm{Jun} \sur{Li}}\email{junli@cma.gov.cn}

\affil[1]{\orgdiv{Key Laboratory of Polar Atmosphere-Ocean-Ice System for Weather and Climate of Ministry of Education/ Shanghai Key Laboratory of Ocean-Land-Atmosphere Boundary Dynamics and Climate Change, Department of Atmospheric and Oceanic Sciences \& Institutes of Atmospheric Sciences}, \orgname{Fudan University}, \orgaddress{\city{Shanghai}, \postcode{200438}, \country{China}}}
\affil[2]{\orgdiv{Chinese Academy of Meteorological
Sciences‐Jiangsu Meteorological Service}, \orgname{Nanjing Joint Institute for Atmospheric Sciences}, \orgaddress{\city{Nanjing}, \postcode{210041}, \country{China}}}
\affil[3]{\orgname{Jiangsu Key Laboratory of Severe Storm Disaster Risk / Key Laboratory of Transportation Meteorology of CMA}, \orgaddress{\city{Nanjing}, \postcode{210041}, \country{China}}}
\affil[4]{\orgdiv{RIKEN Interdisciplinary Theoretical and Mathematical Sciences (iTHEMS)}, \orgname{RIKEN}, \orgaddress{\city{Wako, Saitama}, \postcode{351-0198}, \country{Japan}}}
\affil[5]{\orgdiv{Institute of Atmospheric Physics}, \orgname{Chinese Academy of Sciences}, \orgaddress{\city{Beijing}, \postcode{100029}, \country{China}}}
\affil[6]{\orgdiv{School of Atmospheric Sciences}, \orgname{Nanjing University}, \orgaddress{\city{Nanjing}, \postcode{210023}, \country{China}}}
\affil[7]{\orgdiv{Engineering Research Center of Optical Instrument and System, the Ministry of Education, Shanghai Key Laboratory of Modern Optical System}, \orgname{University of Shanghai for Science and Technology}, \orgaddress{\city{Shanghai}, \postcode{200093}, \country{China}}}
\affil[8]{\orgdiv{National Satellite Meteorological Center}, \orgname{China Meteorological Administration}, \orgaddress{\city{Beijing}, \postcode{100081}, \country{China}}}

\newgeometry{top=2.5cm, bottom=2.5cm, left=3cm, right=3cm}

\abstract{Clouds play a critical role in Earth’s hydrological and energy cycles, and accurately representing their properties is essential for effective numerical modeling and weather forecasting. Machine learning methods have been widely used for cloud property retrieval; however, most existing techniques are deterministic and do not incorporate uncertainty quantification. Generative machine learning has made significant advances in various domains, including natural language processing, image generation, and notably weather forecasting, where it has enabled ensemble predictions and the quantification of forecast uncertainty. This ability to quantify uncertainty offers valuable opportunities for cloud remote sensing. In this study, we propose a novel cloud property retrieval method, CloudDiff, based on a generative diffusion model. By leveraging thermal infrared observations from the Himawari-8 Advanced Himawari Imager (AHI), CloudDiff generates high spatiotemporal resolution cloud properties for both daytime and nighttime conditions, increasing the resolution of Himawari-8/AHI cloud retrievals from 2 km to 1 km. Unlike deterministic retrieval methods, CloudDiff generates multiple samples from the underlying probability distribution, allowing for a diverse range of plausible retrievals and taking steps towards providing uncertainty assessment. Additionally, CloudDiff produces sharper samples and better captures fine local features, enhancing the precision of cloud property retrieval. By averaging over the ensemble of generated samples, we demonstrate that both the accuracy and reliability of the retrievals are significantly improved. These high-resolution cloud properties have been successfully applied to analyze extreme weather events, such as typhoons, providing potentially valuable insights into atmospheric processes.}





\maketitle
\section{Introduction}\label{sec1}
Recent years have witnessed remarkable advancements in generative machine learning, especially in image generation. For example, OpenAI's DALL-E 3 \citep{betker2023} and Stability AI's Stable Diffusion \citep{rombach2022} utilize conditional sampling mechanisms based on text prompts to generate diverse visual outputs through probabilistic modeling of high-dimensional data distributions. These models produce non-deterministic, diverse images that effectively capture the full spectrum of user requirements, offering insights into the handling of uncertainty in complex systems, such as those in atmospheric science. Specifically, in weather forecasting, the inherent complexity and nonlinearity of atmospheric processes present substantial challenges in predicting future weather states. Traditional deterministic forecasting methods often fail to account for the full range of possible scenarios, resulting in significant uncertainties in predictions \citep{li2024}. In contrast, generative machine learning constructs ensemble forecasts that provide probabilistic representations of plausible atmospheric states. This approach not only quantifies uncertainty but also estimates the likelihood of extreme events, thereby improving predictive accuracy \citep{price2025}. The shift from deterministic predictions to probabilistic ensemble sampling represents a paradigm shift in uncertainty quantification within atmospheric science.

Clouds play a vital role in the atmospheric system \citep{stephens1990, li2005} and are the primary source of uncertainty in studies of weather, climate change, and climate modeling\citep{heintzenberg2009, zelinka2017}. They regulate the global energy balance and influence the water cycle \citep{wang2016a, fauchez2018, min2020}, and are strongly coupled with aerosols, involving complex feedback mechanisms across multiple temporal and spatial scales \citep{stevens2009, chen2018}. Cloud properties, such as cloud optical thickness (COT), cloud effective radius (CER), cloud top height (CTH), and cloud phase (CLP), are key parameters that reflect both the macro- and microphysical characteristics of cloud systems, profoundly impacting Earth's net radiation balance \citep{leinonen2019, li2013}. For example, the COT and CER influence clouds ability to reflect and absorb radiation \citep{wang2022}, thereby shaping its net radiative effect and influencing weather and climate. Therefore, acquiring accurate cloud properties is crucial for improving the understanding of cloud evolution under different thermodynamic conditions, reducing uncertainties in cloud simulations within climate models, and understanding their significant radiative effects on weather and climate change.

Satellite remote sensing technology (including that on board geostationary and polar orbiting satellites) is widely used in cloud observation and cloud properties retrieval\citep{king1992, platnick2003, menzel2008, zhang2011, tang2017, zhuge2020, wang2022o}. Geostationary satellites, such as Fengyun-4A and Himawari-8, provide extensive observational coverage and continuous monitoring, enabling the creation of long-term cloud observation time series. Consequently, geostationary satellites have been widely used for cloud property retrieval \citep{wang2022, tong2023, zhao2023, li2023}. Current retrieval methods based on geostationary satellite data primarily include physical methods \citep{nakajima1990, toshiro1985, parol1991, rodgers2000, wang2016a, wang2016b, iwabuchi2016} and machine learning-based approaches \citep{wang2022o, zhao2023}. Machine learning methods have gained widespread adoption for cloud property retrieval due to their ability to automatically extract cloud structural features, perform all-day retrievals using thermal infrared (TIR) observations, and maintain high computational efficiency.  However, current machine learning methods are deterministic, limiting their ability to capture uncertainty, particularly in extreme weather scenarios where satellites observational noise and model errors can significantly amplify uncertainties. The quantification of these uncertainties has become an increasingly important issue. For instance, accounting for uncertainty in the assimilation of cloud properties can significantly enhance the accuracy of simulations for extreme weather events, such as heavy precipitation and typhoons \citep{chen2015, jones2015, meng2022}, as well as for meso- and micro-scale systems \citep{emanuel1994, miyamoto2013, demeutter2015}.

Diffusion models, a class of generative models based on maximum likelihood estimation through score matching, have recently gained attention for their success in generating high-quality images \citep{sohl2015, song2019}. These models offer several advantages, including the ability to capture full probability distributions \citep{ho2020}. Unlike deterministic methods, diffusion models can approach the underlying probability distributions and sample from the target distribution by gradually denoising a reference Gaussian distribution \citep{ho2020, bishop2024, ling2024}, thereby generating a range of plausible samples. This enables the retrieval of a distribution of outcomes, offering insights into the probability density and its score. Diffusion models have shown remarkable success in diverse fields, such as image generation and super-resolution in computer vision \citep{croitoru2023}, precipitation nowcasting \citep{Asperti2025}, estimation of unresolved geophysical processes \citep{pan2023}, and downscaling Earth system models \citep{hess2024}.  Thus, diffusion models offer a promising new approach to retrieving cloud properties by facilitating ensemble sampling and probabilistic representation. However, their potential applications in cloud remote sensing remain largely unexplored.

In this study, we introduce a novel generative diffusion model framework for retrieving cloud properties (including COT, CER, CTH, and CLP) using geostationary meteorological satellite observations. This framework overcomes the limitations of existing methods by enabling high-resolution, all-day cloud property ensemble retrievals and quantifying uncertainties. Unlike previous studies, our approach uses ensemble retrieval techniques based on a diffusion model, capable of identifying CLP and retrieving cloud properties with high accuracy and spatial resolution (1 km). To the best of our knowledge, this is the first cloud property retrieval model using geostationary satellites to achieve high-resolution ensemble retrievals across all-day. By employing probabilistic sampling, the model generates an ensemble of retrieval outcomes, providing a comprehensive characterization of uncertainty, accounting for model retrieval errors. This study not only extends the application of deep generative model techniques in cloud properties retrieval but also presents a novel methodology that can be adapted for the retrieval of other atmospheric variables from satellite observations, including temperature, humidity, greenhouse gases, pollutants, and surface radiation.

\section{Results}
The diffusion model for cloud properties retrieval, referred to as CloudDiff, utilizes 2 km thermal infrared observations from Himawari-8/AHI as conditions (i.e., model's inputs), achieving high spatial resolution (1 km) retrievals of cloud properties. The model’s performance is evaluated against the standard cloud product of Moderate Resolution Imaging Spectroradiometer (MODIS), Cloud-Aeroso Lidar and Infrared Pathfinder Satellite Observation (CALIPSO), and Himawari-8, focusing on its generalization capabilities and uncertainty quantification. Additionally, the potential applications of the retrieved cloud properties are demonstrated through an analysis of typhoon In-Fa. Further details on the datasets, model architecture, and evaluation metrics are provided in the Data and Methods section.

\subsection{Evaluation using MODIS cloud products}
CloudDiff can generate multi-sample retrieval results through probabilistic distribution sampling. To evaluate its performance, we conducted a comparative analysis using MODIS cloud products as reference data. Meanwhile, a comparative analysis was performed to examine the impact of sample size on high-resolution retrieval performance, as well as to understand the advantages of ensemble retrieval. The analysis included a single sample and ensemble means calculated from 5 and 30 samples, which were compared against the results from the regular UNet model \citep{trebing2021} (hereafter referred to as the deterministic model).

\begin{figure*}[hb]
\centering
\includegraphics[width=1\textwidth]{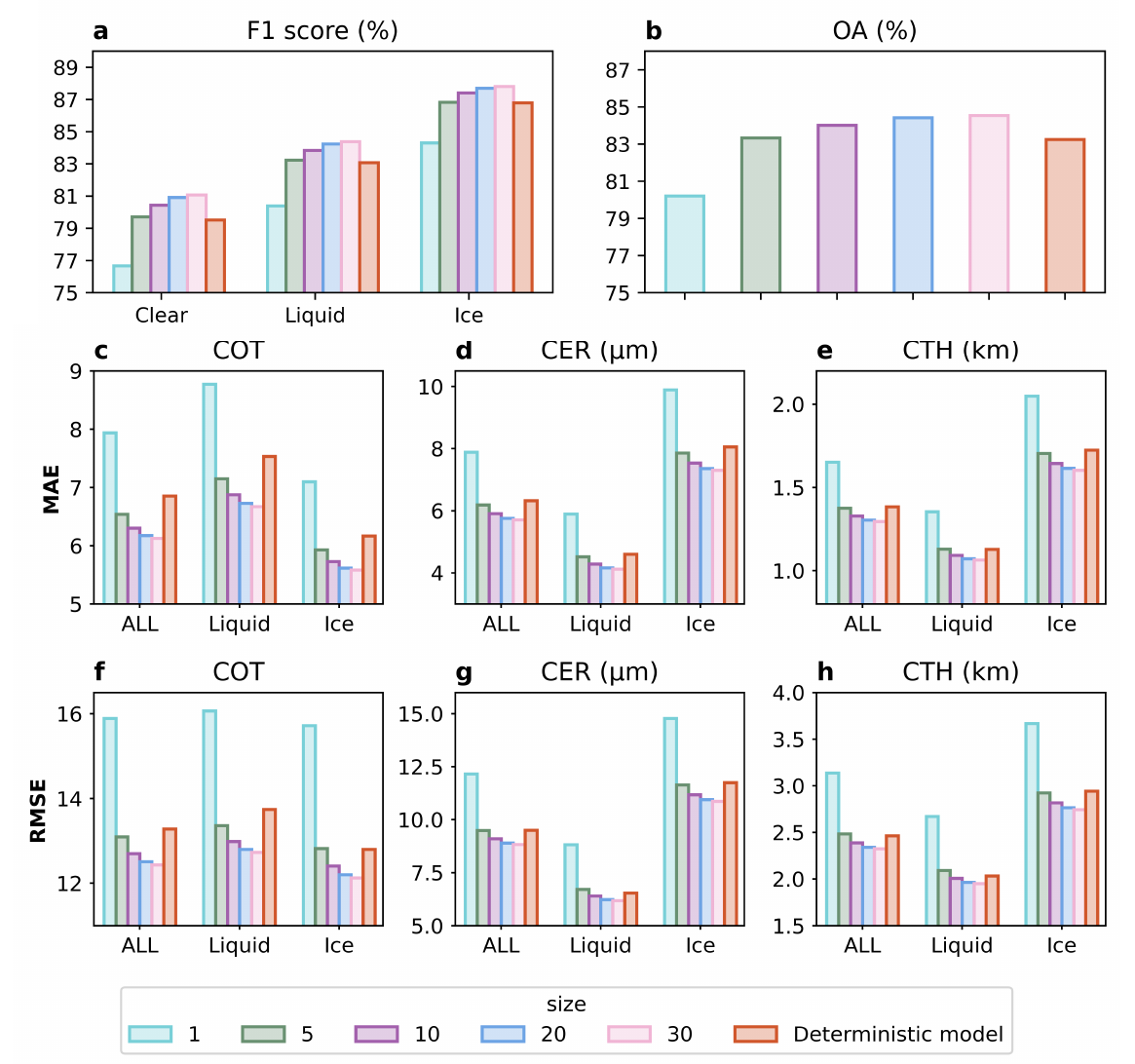}
\caption{The performance evaluation of retrieved CLP and cloud properties. The bars, colored differently, represent sample sizes ranging from 1 to 30. (a) displays the F1 scores for different CLPs, comparing single samples and ensemble means of various sample sizes (ranging from 5 to 30) with the deterministic model; (b) shows the  overall accuracy (OA) for all CLPs, comparing single samples and ensemble means of various sample sizes with the deterministic model; (c--e) and (f--g) show the MAE and RMSE for cloud properties, respectively, comparing single samples and ensemble means of various sample sizes with the deterministic model.}\label{fig1}
\end{figure*}

We first evaluated the results of CLP identification, which classifies the targets into clear sky, liquid clouds, and ice clouds. To assess the performance, we compared the deterministic model with the ensemble mean of CloudDiff, calculated from varying sample sizes (Fig. \ref{fig1}a,b). For a single sample, the F1 score and overall accuracy (OA) of CLP identification is the lowest. As the ensemble size increases, both the F1 score and OA of the ensemble mean gradually increase, reaching the highest accuracy at an ensemble size of 30. CloudDiff achieves an OA of 84.55\% (Fig. \ref{fig1}b), with retrieval accuracies of 85.13\% for clear sky, 82.66\% for liquid clouds, and 87.21\% for ice clouds (Fig. S1 in Supplementary Material). In contrast, the deterministic model achieves comparable accuracies for CLP identification to those obtained with an ensemble size of 5, with 86.09\% for ice clouds and 81.62\% for liquid clouds, resulting in an overall accuracy (OA) of 83.25\%, which is slightly lower than that of CloudDiff. Both CloudDiff and the deterministic model tend to confuse clear sky with liquid clouds, as well as liquid clouds with ice clouds (Fig. S1 in Supplementary Material). Despite some errors, the ensemble mean of CloudDiff demonstrates superior performance in retrieving cloud properties and identifying clear sky, liquid clouds, and ice clouds compared to the deterministic model. 

Additionally, the performance on cloud properties was evaluated. The comparison of Mean Absolute Error (MAE) and Root Mean Squared Error (RMSE) between the MODIS cloud products and the retrieved results is also shown in Figure \ref{fig1}c--h. Similar to the results of CLP identification, the comparison reveals significantly higher MAE and RMSE values when using a single sample. As the ensemble size increases beyond five, both the MAE and RMSE of the ensemble mean gradually decrease. Interestingly, the improvement in high-resolution retrieval capability from 20 to 30 samples is relatively minor, suggesting that approximately 30 samples are sufficient to capture most high-resolution details and adequately cover the uncertainty space in the retrieval process. The MAE and RMSE values of the deterministic model approach those obtained with an ensemble size of 5 and are notably higher than those observed with an ensemble size of 30.

Specifically, for COT at an ensemble size of 30, the ensemble mean MAE for all clouds (liquid and ice) is 6.12, with an RMSE of 12.43. In comparison, the deterministic model results have an MAE of 6.85 and an RMSE of 13.28, with ice clouds showing slightly better performance. For CER, the ensemble mean MAE for all clouds at an ensemble size of 30 is 5.70 $\mu$m, with an RMSE of 8.83 $\mu$m, and liquid clouds exhibit lower MAE and RMSE than ice clouds. In contrast, the deterministic model's MAE is 6.32 $\mu$m and RMSE is 9.51 $\mu$m, both higher than those of CloudDiff. Similarly, for CTH at the same ensemble size, the ensemble mean MAE for all clouds is 1.29 km, with an RMSE of 2.32 km, and ice clouds perform worse than liquid clouds. This difference performance in liquid and ice clouds is closely tied to their characteristics and intrinsic values. For example, liquid clouds predominantly consist of tiny water droplets (approximately 10 $\mu$m) and are generally found in the lower atmosphere, where temperatures are relatively warmer. In contrast, ice clouds are primarily composed of ice crystals and are typically located in the upper atmosphere, where temperatures are much colder. Therefore, in general, the CER and CTH values of liquid clouds are smaller than those of ice clouds. As a result, for the CloudDiff algorithm, the RMSE for CER and CTH in liquid clouds is smaller than that in ice clouds for the same percentage error.

Ensemble mean RMSE and ensemble spread (i.e., standard deviation) are widely used metrics for assessing the reliability of ensemble prediction systems. In a well-designed system, the spread among ensemble members should approximately match the ensemble mean RMSE, thereby ensuring a high probability of capturing the true cloud properties even with a limited number of samples. Accordingly, we compared the ensemble spread and ensemble RMSE (Fig. S2 in the Supplementary Material). The results show that across different values of COT, CER, and CTH, the ensemble spread and ensemble RMSE are similar in magnitude, but the spread is generally smaller, indicating the presence of systematic biases in CloudDiff. To  examine the spatial distribution of these errors, we evaluated CloudDiff across viewing zenith angles (VZA; 0–72°) in the test set (Fig. S3 in the Supplementary Material) and compared the results with those of the deterministic model. For CloudDiff, COT retrievals are sensitive to viewing geometry: as the satellite’s nadir is near the equator, increasing VZA enlarges the ground-projected pixel area, reducing spatial resolution and retrieval accuracy. Errors in CER and CTH depend not only on VZA but also on the latitudinal distribution of cloud properties, with higher values near the equator and lower values at higher latitudes \citep{chi2024}. For CLP, overall accuracy and F1 scores peak at small VZAs, decline at 10–20°, and vary by ~1\% thereafter. Compared with CloudDiff, the deterministic model shows similar VZA-dependent trends but consistently larger errors.

In addition to conventional cloud property evaluation metrics, we further employed the Peak Signal-to-Noise Ratio (PSNR) and Structural Similarity Index Measure (SSIM) to assess the similarity and quality of the retrieved cloud properties relative to MODIS cloud products (Fig. 2a, b). The results are consistent with those obtained from the MAE and RMSE evaluations. For individual samples, both SSIM and PSNR values are the lowest. As the ensemble size increases, both metrics improve, reaching their maximum at an ensemble size of 30, indicating that the CloudDiff ensemble best approximates the MODIS products. Retrievals from the deterministic model perform comparably to those from an ensemble size of five. Notably, CloudDiff achieves strong structural similarity (SSIM = 0.68) and good pixel-level accuracy (PSNR = 25.87) for COT retrievals, while showing moderate performance for CTH and CER. These comprehensive metrics collectively confirm that ensemble retrieval approach effectively captures both the structural and pixel-level characteristics of MODIS cloud products.

Additionally, we computed the power spectral density (PSD) of cloud properties as a function of spatial wavelength (Fig. 2c–e) to evaluate the physical realism of the CloudDiff retrievals. For COT and CER, a single sample captures the spatial variability with reasonable fidelity, whereas for CTH, a single sample reproduces the large-scale patterns ($>5$ km) but fails to retain realism at smaller scales ($<5$ km). Increasing the ensemble size progressively smooths small-scale structures, and at an ensemble size of 30 the ensemble mean fails to represent both large- and small-scale features, as averaging inevitably blurs the sharpness of individual samples. Nevertheless, ensemble means still outperform the deterministic model, which poorly represents cloud structures. Overall, the single ensemble sample provides sharper representations and better preserves realism across scales (particularly for COT), whereas the ensemble mean smooths fine-scale details but still outperforms deterministic retrievals.

\begin{figure}[H]
\centering
\includegraphics[width=1\textwidth]{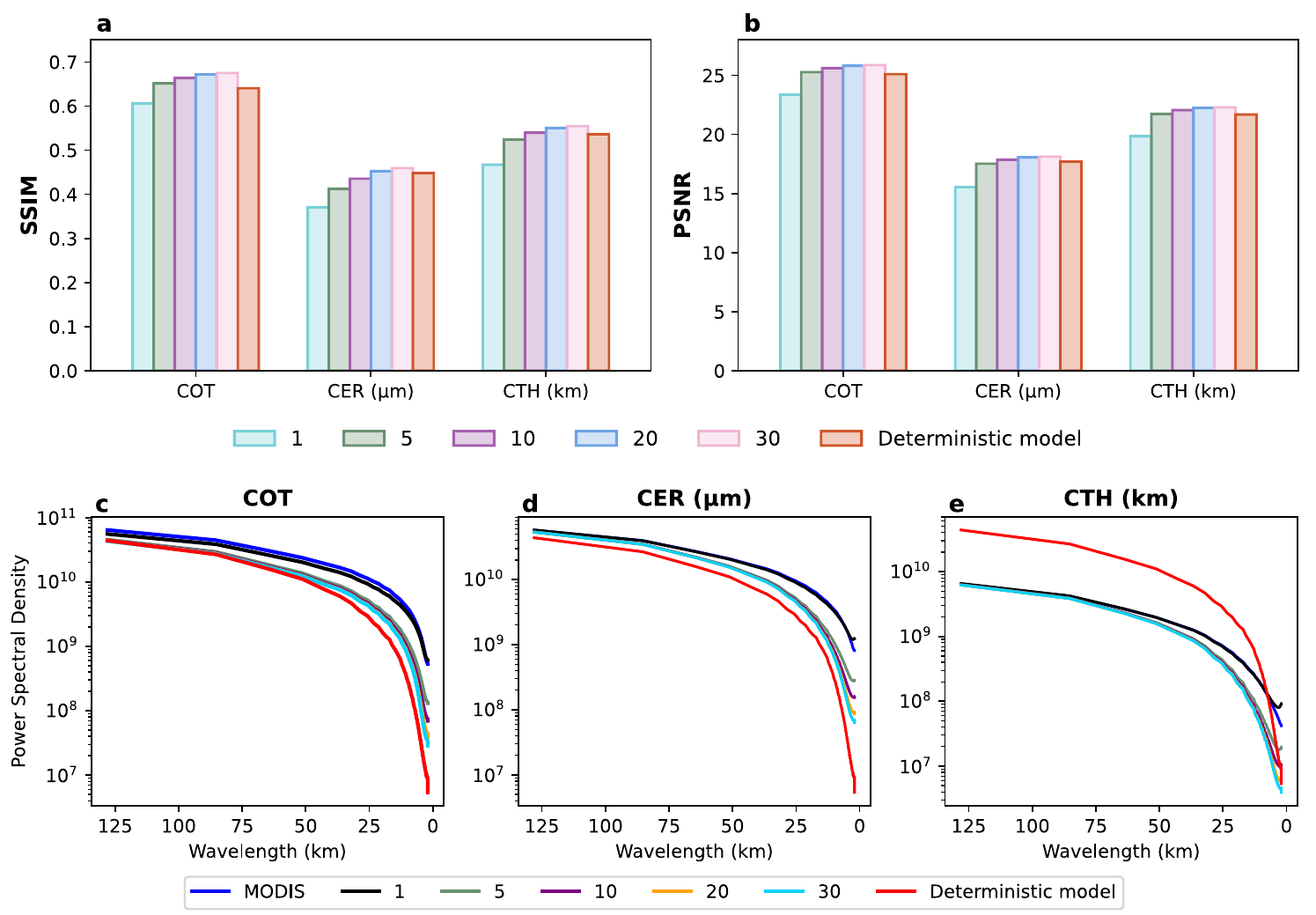}
\caption{Performance evaluation and power spectrum analysis of the retrieved cloud properties. (a) and (b) show the SSIM and PSNR of the cloud properties, respectively, comparing ensemble means of various sample sizes with the deterministic model. (c–e) display the radially averaged power spectra of COT, CER, and CTH, comparing CloudDiff and the deterministic model against MODIS.}\label{fig2}
\end{figure}

Overall, the ensemble mean of multiple samples effectively reduces retrieval errors and improves retrieval accuracy. However, ensemble-averaged images tend to be blurred and represent cloud structures less realistically than a single sample. If higher-fidelity cloud properties are required, single-sample results can be used, albeit at the cost of reduced accuracy. Notably, although averaging 30 ensemble samples yields good retrieval accuracy in this study, it remains uncertain whether the ensemble mean adequately represents the posterior distribution. To address this, we compared retrieval errors obtained using the ensemble median and the maximum a posteriori estimate based on kernel density estimation (Table S1 in the Supplementary Material). Both alternatives resulted in slightly larger errors compared to the ensemble mean. Thus, these findings indicate that the ensemble mean serves as a reasonable descriptor of the posterior distribution. Besides, since CloudDiff retrieval errors stabilize at an ensemble size of 30, increasing the ensemble size further is unlikely to substantially improve performance while incurring additional computational cost. Therefore, for subsequent analyses, the ensemble mean is computed using 30 samples.

\subsection{Evaluation using other cloud products and uncertainty analysis}
To further validate CloudDiff, we performed a comprehensive evaluation by comparing its retrievals with active sensor measurements from the CALIPSO lidar Level-2 products (CTH and CLP). Operational CTH and CLP retrievals from Himawari-8 were also included in the assessment. Specifically, we compared CloudDiff, the deterministic model, Himawari-8, and MODIS against CALIPSO (Fig. 3). The results show that CloudDiff’s CTH achieves the highest agreement with CALIPSO, with a MAE of 1.73 km and an RMSE of 2.81 km, slightly outperforming both the deterministic model and Himawari-8, and performing comparably to MODIS. Nevertheless, all approaches exhibit varying degrees of CTH underestimation for clouds above 10 km. The magnitude of underestimation in CloudDiff closely matches that of MODIS, indicating that while CloudDiff has effectively learned the MODIS CTH distribution, it has also inherited its biases.

For CLP classification, CloudDiff achieves an OA of 83.44\% and consistently outperforms the deterministic model across all cloud phase categories. The operational product from Himawari-8 performs better than CloudDiff for liquid clouds but is notably less accurate for clear-sky and ice cloud identification when validated against CALIPSO, leading to lower overall performance. Since CloudDiff was trained on MODIS data, its CLP classification closely aligns with MODIS, exhibiting similar errors across all cloud phase categories. Overall, CloudDiff’s CTH and CLP retrievals are reliable, showing strong agreement with CALIPSO lidar observations, while inevitably inheriting the biases present in MODIS.

\begin{figure}[H]
\centering
\includegraphics[width=1\textwidth]{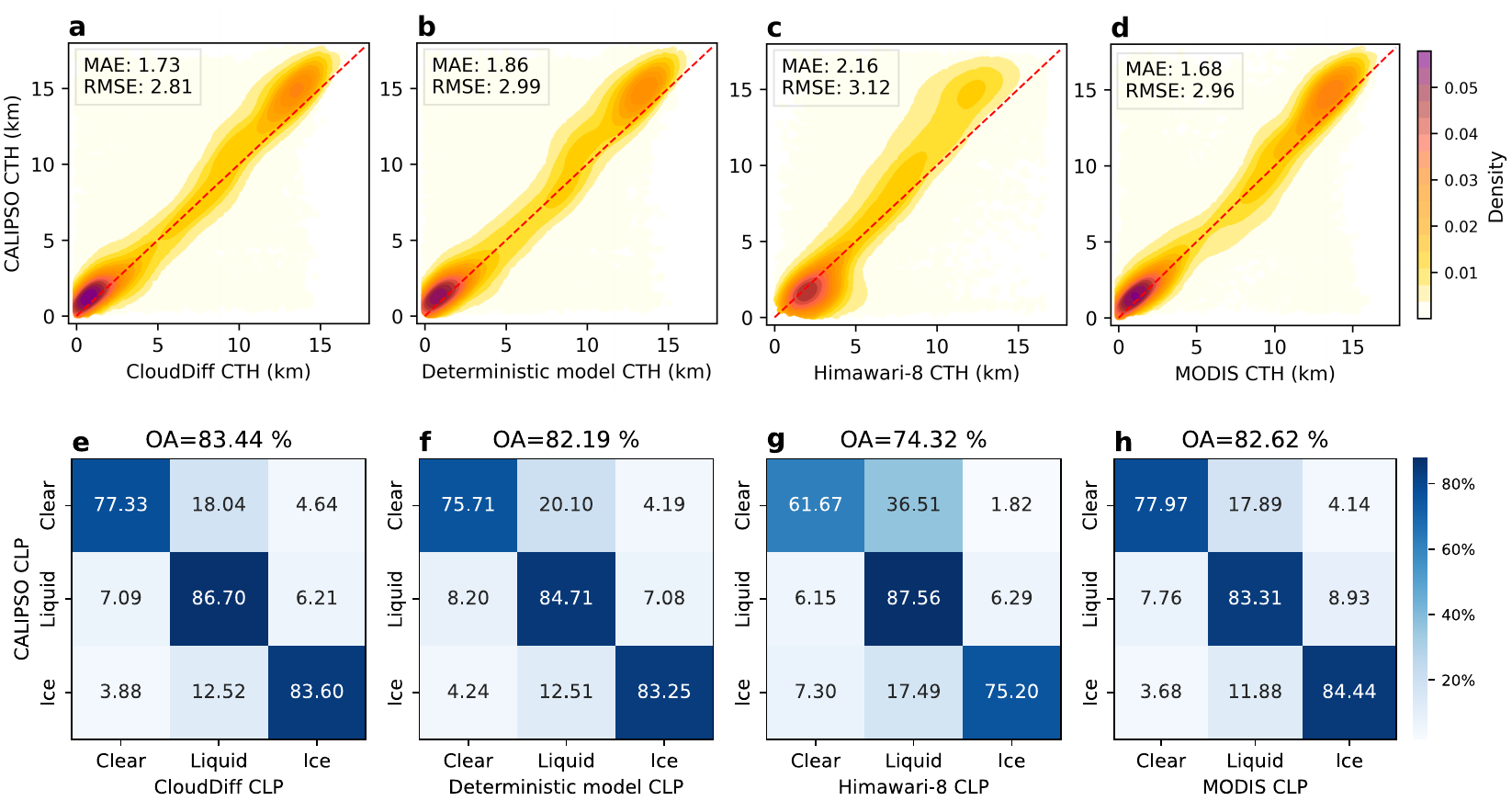}
\caption{Joint probability density plots of CTH (first row) and confusion matrices of CLP (second row). (a–d) show the joint probability density distributions of CALIPSO with (a) CloudDiff, (b) the deterministic model, (c) Himawari-8, and (d) MODIS. (e–h) present confusion matrices comparing CALIPSO CLP products with (e) CloudDiff, (f) the deterministic model, (g) Himawari-8, and (h) MODIS.}\label{fig3}
\end{figure}

In weather forecasting, ensemble spread is commonly used as a metric to quantify the uncertainty of ensemble predictions relative to the ensemble mean. Similarly, we use ensemble spread, expressed as the standard deviation, to represent retrieval uncertainty. This choice is partly motivated by the fact that the retrieval errors based on the ensemble mean, ensemble median, and maximum a posteriori estimate from kernel density estimation are relatively close in magnitude. This suggests that the sample distributions generated by CloudDiff are close to Gaussian. As shown above, the CTH retrieved by CloudDiff exhibit certain errors and inevitably reproduce biases present in MODIS. This raises the critical question of whether the estimated uncertainty reflects only model-related uncertainties in the retrieval process, or also encompasses the inherent biases of MODIS.

To address this question, we further analyzed the bias distributions of MODIS and CloudDiff relative to CALIPSO, as well as the changes in bias when uncertainties are considered (Fig. S4 in the Supplementary Material). The results indicate that MODIS biases relative to CALIPSO approximately follow a normal distribution and generally underestimate CTH. CloudDiff reproduces these biases, and its deviations relative to CALIPSO also follow a normal distribution, showing overall underestimation slightly greater than that of MODIS. Incorporating retrieval uncertainties into CloudDiff (Fig. S4d in the Supplementary Material) partially alleviates the underestimation (upper bound of uncertainty) but does not fully correct it. This suggests that CloudDiff’s uncertainty (i.e., standard deviation) captures retrieval-related errors but does not account for intrinsic MODIS biases, leading to somewhat underestimated uncertainties.

\subsection{Application to typhoon case}
As demonstrated above, by leveraging the ensemble mean retrieval results from CloudDiff, accurate cloud properties (1 km, 10 min) for all-day retrieval can be obtained. These cloud properties have potential applications across various fields, particularly in monitoring the evolution of extreme weather, which inherently involves uncertainties. To illustrate this, we applied the cloud properties generated by CloudDiff in a case study of typhoon In-Fa (No. 2106), further validating the ensemble retrieval performance of CloudDiff, while gaining an understanding of the associated uncertainties and exploring its potential applications.

Typhoon In-Fa, the sixth typhoon of 2021, formed over the Northwest Pacific Ocean at 18:00 UTC on July 17. Analysis of the best track and intensity data of typhoon In-Fa (Fig. S5 in Supplementary Material) shows that by 00:00 UTC on July 21, “In-Fa” had intensified to a strong typhoon (42 m/s) and subsequently altered its trajectory from westward to northwestward. By 00:00 UTC on July 23, it weakened to a typhoon level as it approached the Zhejiang coast. “In-Fa” made landfall in the coastal area of Zhoushan, Zhejiang Province, China, around 04:30 UTC on July 25. By 12:00 UTC on July 30, it had weakened further, eventually transitioning into an extratropical cyclone over the Bohai Sea.

\begin{figure}[H]
\centering
\includegraphics[width=1\textwidth]{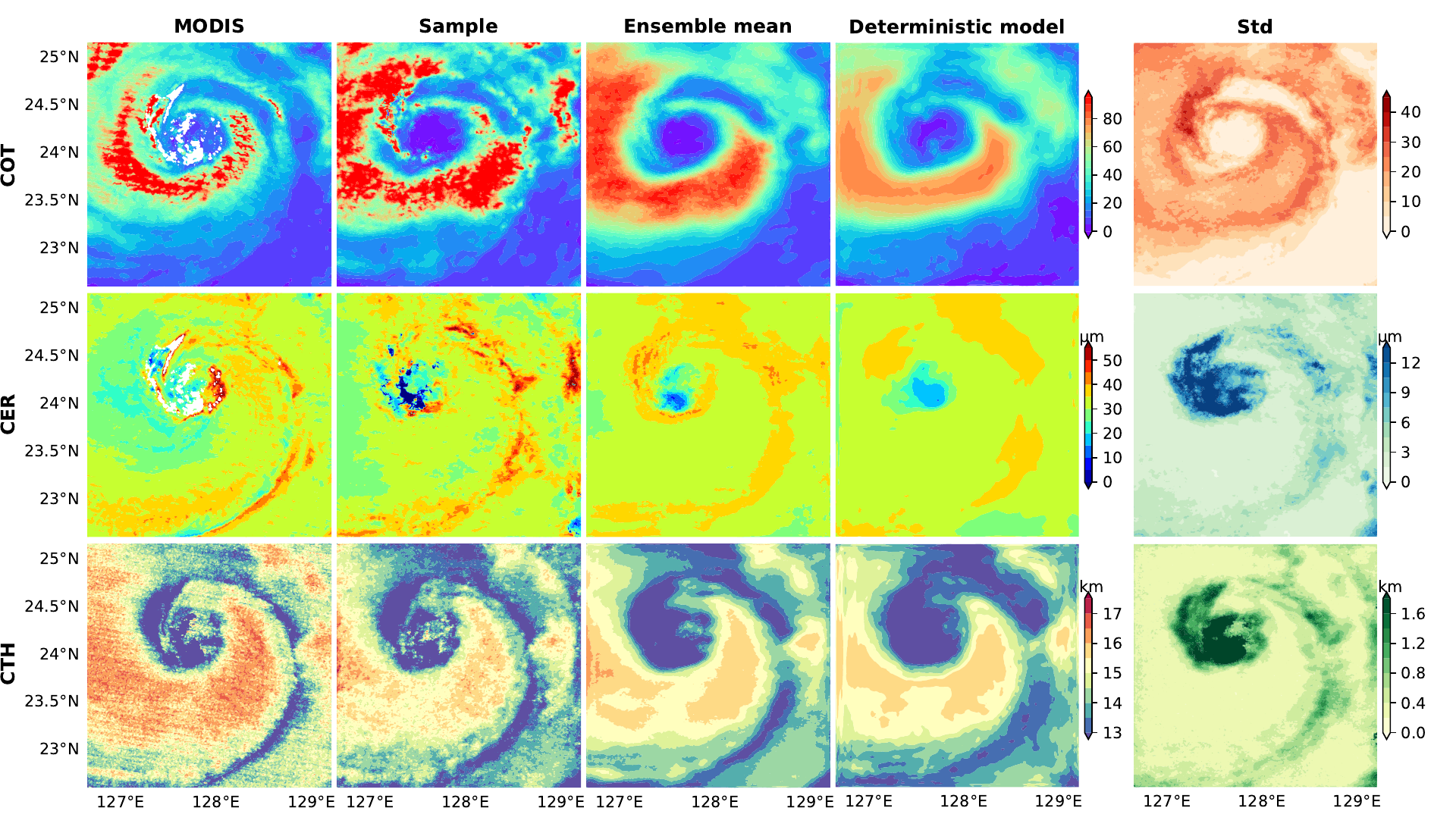}
\caption{MODIS cloud products and retrieved cloud properties in the typhoon In-Fa region, centered at 24.1°N, 127.8°E, at 0220 UTC on July 21, 2021. The columns are MODIS cloud products, sample, ensemble mean, deterministic model, and standard deviation (std). Note the blank areas in the MODIS cloud products represent missing data.}\label{fig4}
\end{figure}

Here, we focus on the typhoon’s condition when it reached a strong typhoon intensity on July 21. On that day, the Terra satellite passed over the East China Sea region at approximately 02:20 UTC; therefore, our analysis primarily focused on evaluating the retrieved cloud properties at this specific time. The analysis covered the typhoon area between 22.61°N–25.17°N and 126.65°E–129.21°E. To evaluate the performance and uncertainty of CloudDiff, we compared the cloud properties with the MODIS cloud products, the results from the deterministic model are also displayed (Fig. \ref{fig4}). Clearly, CloudDiff's retrieval results outperform those of the deterministic model, particularly for extreme values. Although previous assessments indicate that the ensemble mean has lower error than a single sample, the results show that a single sample produces sharper images and more local details of COT, CER, and CTH compared to the ensemble mean (which appears blurrier), averaged over 30 samples. The deterministic model’s results are even blurrier than the ensemble mean and lack detail. The results are consistent with the previous power spectrum analysis.

Specifically, regarding COT, compared to MODIS cloud products, the single sample underestimated COT in the typhoon eye region and overestimated it in areas where COT\textless 90. The ensemble mean similarly overestimated the extent of areas with COT\textless 90 but reported lower values than the single sample, somewhat mitigating the overestimation of COT in the typhoon eyewall region by the single sample. The standard deviation of the 30 samples, representing retrieval uncertainty, reveals substantial COT uncertainty within the typhoon’s eyewall. However, the deterministic model not only underestimated the extent of areas with COT\textgreater 90 but also presented an overestimation on the northeastern side of the typhoon eye. Both the single sample, ensemble mean, and deterministic model inaccurately retrieved areas with CER\textgreater 40 $\mu$m. However, CloudDiff exhibited smaller biases in CER retrievals compared to the deterministic model, with standard deviations mostly below 6 $\mu$m across most regions, indicating small uncertainty except for the typhoon eye region.

Regarding CTH, CloudDiff exhibits low uncertainty, with standard deviations typically below 1.5 km across most regions. Compared to MODIS, a single sample appears to better represent the CTH distribution, while the ensemble mean tends to underestimate regions with CTH\textgreater 15 km, particularly in the typhoon eye region. The deterministic model also underestimates CTH\textgreater 15 km, as well as CTH in the typhoon eye and at the image edges. Overall, CloudDiff's retrieval of cloud properties outperforms the deterministic model, providing sharper, more localized details of 1 km cloud properties during the typhoon event. It is worth noting that when cloud property values are small, high uncertainty may lead to unphysical (negative) values. To evaluate the appropriateness of the uncertainty, we examined the range of mean ± standard deviation and calculated the proportion of negative values for COT, CER, and CTH, which were 1.57\%, 0.19\%, and 0.18\%, respectively. Therefore, although the standard deviation may occasionally include unphysical values, their occurrence is minimal, indicating that most retrievals remain physically reasonable. Future research will adopt a physics-driven deep learning approach \citep{aarts2025} to update the CloudDiff model, imposing rigorous physical constraints on the neural network’s outputs. For example, positive-definite activation functions (e.g., Softplus) may be incorporated to ensure the strict positivity of cloud property predictions. Additionally, it is important to note that CLP represents different cloud phases as categorical values. Therefore, quantitative standard deviation was not provided. 

\begin{figure}[H]
\centering
\includegraphics[width=1\textwidth]{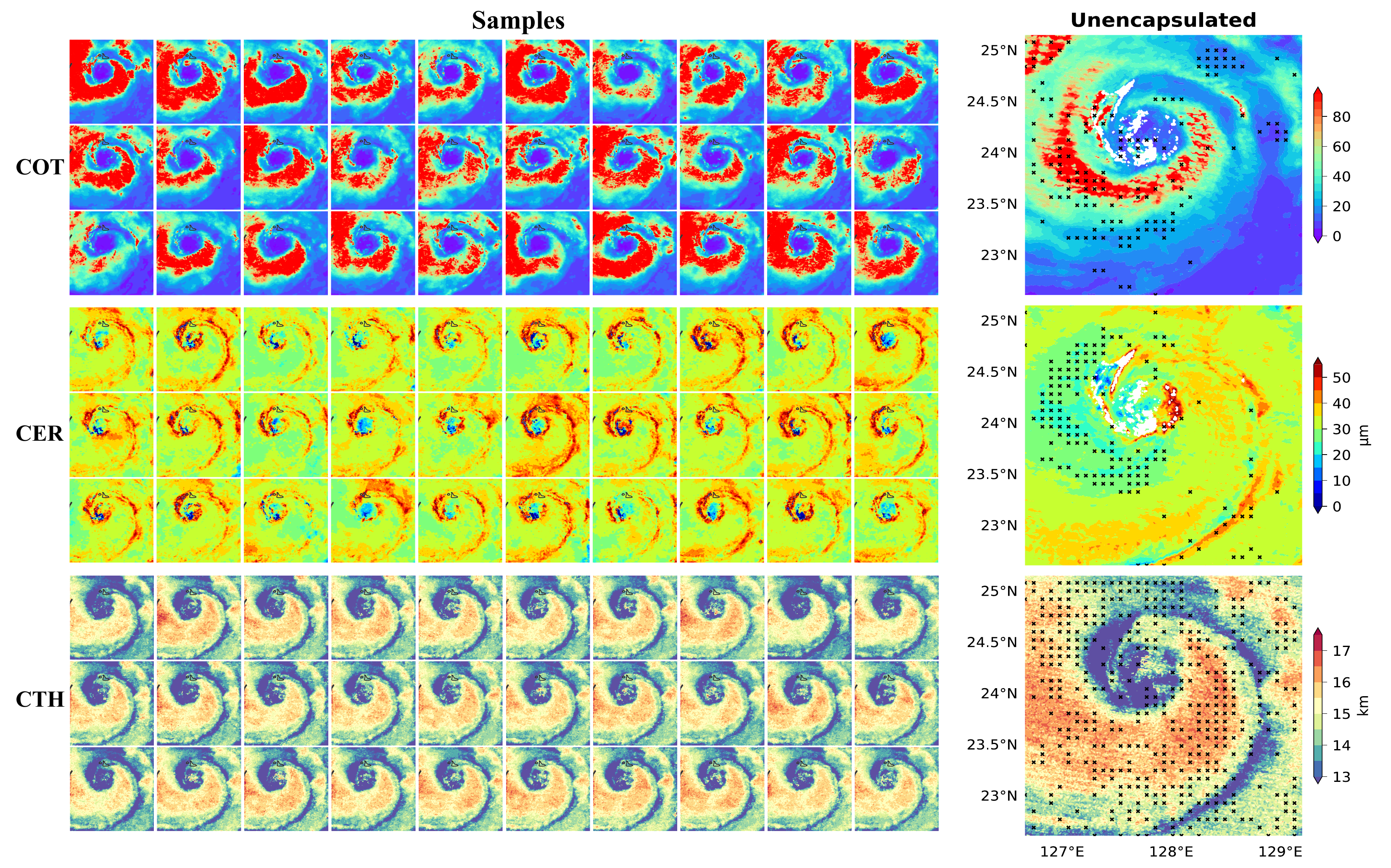}
\caption{The retrieval of cloud properties in the typhoon In-Fa region at 02:20 UTC on July 21, 2021, was performed using the CloudDiff. The columns represent different samples and grid points where MODIS cloud properties are not captured by the samples. Underestimations and overestimations are indicated by black ’x’ markers. The background is colored based on MODIS cloud products, with blank areas indicating missing data.}\label{fig5}
\end{figure}

As shown above, although the ensemble mean can reduce retrieval errors, individual samples produce sharper images and capture more local details, providing valuable information, much like ensemble members in numerical weather prediction. Therefore, we further present the samples of the retrieved cloud properties. Figure \ref{fig5} displays the cloud properties generated by CloudDiff across 30 samples, along with grid points where MODIS cloud properties were not captured. The distribution of COT values varies among the samples; underestimation occurred in approximately 6.41\% of the grid points, while overestimation occurred in 8.58\% of the grid points with COT values. Nevertheless, some samples successfully captured high COT values in the typhoon eyewall region and low COT values in the typhoon eye. Regarding CER, some samples accurately reflected the variability in CER values, although some samples exhibited overestimations or underestimations, generally overestimating it in 11.36\% of grid points and underestimating it in only 2.09\% of grid points. Additionally, CTH was underestimated to varying degrees, particularly across the entire study region, with a total underestimation of 34.46\% and an overestimation of only 4.34\%. However, some samples were able to accurately estimate high CTH values in the typhoon eye and the western part of the typhoon. Due to differences among the CLP samples, all cloud phases were effectively captured (not shown). The results indicate that CloudDiff enables the efficient generation of realistic samples that align well with a wide range of retrieval schemes while remaining sufficiently diverse to encompass most plausible outcomes. Overall, these high spatiotemporal resolution, all-day cloud properties demonstrate considerable potential for identifying and monitoring the evolutionary processes of typhoons.

\section{Discussion}
Here, we propose a conditional diffusion model named CloudDiff for the high-resolution retrieval of COT, CER, and CTH, as well as CLP identification. The model is trained on TIR measurements from AHI onboard Himawari-8, using MODIS cloud products as target data. CloudDiff is capable of generating cloud properties and CLP with high resolution (1 km). It can produce various samples to effectively cover the distribution of most cloud property values, including some with lower occurrence probabilities, such as extreme values, and makes steps towards providing uncertainty assessment. This is particularly important because meso/micro-scale fields within a large-scale background are inherently non-deterministic. 

Analysis of the retrieval results revealed that MAE and RMSE decreased with increasing ensemble size, with the relatively small errors observed at an ensemble size of 30. The performance of the deterministic model matches that of the ensemble mean when the ensemble size is 5. These findings underscore that increasing sample size enhances retrieval capabilities, but this improvement is minimal beyond a certain size; for instance, increasing the ensemble size from 20 to 30 offers little improvement. The evaluation demonstrated that CloudDiff performed well in retrieving cloud properties (RMSE: 12.43 for COT, 8.83 $\mu$m for CER, and 2.32 km for CTH) and identifying CLP (OA: 85.13\%). Evaluation of CloudDiff’s performance across different VZAs indicates that retrieval accuracy is influenced not only by observation geometry but also by the distribution of cloud properties, which can vary under different meteorological conditions.

Power spectrum analysis revealed that the ensemble mean fails to capture small-scale cloud properties adequately. Individual samples exhibit greater sharpness than the ensemble mean, enabling better representation of both large- and small-scale structures, especially for COT. This difference likely arises because the ensemble mean tends to blur fine details. Nonetheless, the ensemble mean still offers a more realistic representation of cloud properties than the deterministic model.

Additionally, the results indicate that the ensemble mean appropriately describes the posterior distribution, as CloudDiff-generated cloud properties are approximately Gaussian distribution, and then the standard deviation used to quantify retrieval uncertainty. Comparisons with CALIPSO indicate that CloudDiff achieves high accuracy in CTH and CLP retrievals, showing strong agreement with CALIPSO lidar products and slightly outperforming both the deterministic model and Himawari-8 operational products. While the model successfully captures the distribution of MODIS CTH and CLP, it inevitably inherits MODIS-related biases, which contribute to underestimated uncertainties.

The evaluation of the model in the typhoon In-Fa case demonstrates that the ensemble mean provided by CloudDiff effectively captures the range of COT, CER, and CTH during the typhoon event and accurately identifies CLP. Compared to the deterministic model, CloudDiff’s cloud properties align more closely with MODIS cloud products, enhancing the sharpness of the retrieval results. Additionally, the uncertainty estimates for cloud properties improve the understanding of their range and increase the reliability of the retrievals. It is worth noting that uncertainties may occasionally yield nonphysical (negative) values; however, such instances are rare, and the vast majority of results remain positive definite. The application of these high-resolution cloud properties to typhoon In-Fa underscores their substantial potential in various fields, such as monitoring the development and evolution of typhoons and their associated precipitation.

Although CloudDiff has shown promising results, further improvements are still possible. The MODIS cloud products contain inherent errors and uncertainties, such as a tendency to underestimate CTH, overestimate CER, and underestimate COT. The uncertainties reported by CloudDiff, however, reflect only the model’s retrieval uncertainty and do not account for biases present in MODIS products. Based on its application to the case study of typhoon In-Fa, CloudDiff exhibits a good capacity to learn from extreme/rare events. This may result in some reasonable deviations from the results of CALIPSO products. To mitigate this issue, the model could be fine-tuned using CloudSat or CALIPSO observations in a transfer learning manner, or posterior sampling errors could be reduced through methods such as the Metropolis-Adjusted Langevin Algorithm combined with CALIPSO data. Moreover, because the diffusion model employs the Denoising Diffusion Probabilistic Model sampling procedure—which starts from pure Gaussian noise and iteratively denoises over a predefined number of time steps—inference is relatively time-consuming. With sufficient computational resources, it would be feasible to generate a larger number of samples and identify the optimal ensemble size to further enhance performance. Future work will also involve more studies of high-impact weather events to further assess CloudDiff’s performance and explore specific applications in ensemble retrieval.

We hope that the demonstrated utility of generative artificial intelligence technology for cloud identification and probabilistic retrieval will promote its application in cloud remote sensing, which is crucial for quantifying uncertainty in the identification and forecasting of extreme weather events. Notably, although the high-resolution retrieval method has been applied to Himawari-8/AHI, it is also applicable to other geostationary satellites, including those with coarser resolutions, such as Meteosat/SEVIRI and FY-4A/AGRI. Furthermore, we believe it is now time to explore the potential of diffusion models in related atmospheric fields. Their applicability extends well beyond cloud properties retrieval, offering promising solutions to challenges associated with uncertain events, such as convection forecasting, typhoon forecasting, and cloud data assimilation. 

\section{Methods}
\subsection{Data}
Himawari-8, launched in October 2014, is a geostationary satellite sensor system operated by the Japan Meteorological Agency. It represents the latest iteration in the Multifunctional Transport Satellite series. The AHI sensor onboard Himawari-8 captures full-disk images every 10 minutes across 16 spectral bands from visible to infrared wavelengths, covering regions from East Asia to Australia. The TIR measurements are sensitive to optically thin clouds and are continuously obtained throughout the diurnal cycle, independent of solar geometry \citep{fauchez2018}. In this study, 2 km TIR radiance measurements from Himawari-8 AHI are utilized to estimate cloud properties during both daytime and nighttime. Additionally, the satellite viewing zenith angles (VZA) are employed to construct the retrieval model due to its impact on radiance at different angles.

Since the launch of NASA’s Terra satellite in 1999, followed by Aqua in 2002, MODIS has emerged as one of the most indispensable satellite remote sensing instruments for Earth science research. MODIS measures reflected solar and emitted thermal radiation across 36 spectral channels (0.42–14.24 $\mu$m), offering unique spectral and spatial capabilities for retrieving cloud properties \citep{platnick2016}. The Terra-MODIS (MOD06) and Aqua-MODIS (MYD06) cloud products, which have a spatial resolution of 1 km. These products include cloud top properties (e.g., CTH and CLP for both day and night) and cloud optical and microphysical properties (e.g., COT and CER during daytime only). Over the years, the MODIS cloud products have demonstrated consistently high accuracy and reliable performance \citep{king2003, platnick2015}. Given the lack of nighttime COT and CER products in the MODIS operational products, the daytime MODIS cloud optical and physical properties (CTH, COT, CER, and CLP) from the Level-2 cloud products (MYD06 L2 and MOD06 L2) are utilized as ground truth to develop the high-resolution retrieval model.

Specifically, the TIR measurements (6.95 $\mu$m, 7.35 $\mu$m, 8.60 $\mu$m, 9.63 $\mu$m, 10.45 $\mu$m, 11.20 $\mu$m, 12.35 $\mu$m, and 13.30 $\mu$m) along with the VZA of the Himawari-8 AHI serve as the conditions for the model, while the MODIS Level-2 COT, CER, CTH, and CLP data are used as the targets for training the model. To optimize the model during training and enhance its accuracy, we normalized both conditions and targets. Specifically, we applied min-max normalization to scale all variables to the range of 0 to 1. 

To cover a wide range of viewing geometries and seasonal variations, data from 2016 to 2018 (covering a VZA range of 0°–85°) were collected. Specifically, data from January 2016 to June 2017 were used for model training, data from July to December 2017 for model validation, and data from January, April, July, and October 2018 were used for testing. Due to differences in spatiotemporal resolution between the Himawari-8 AHI and MODIS cloud products, we performed spatiotemporal matching. Data from both MODIS and Himawari-8 were selected for the same regions and times, with the cloud product grid points corresponding to twice the number of TIR observation grid points. To reduce memory and computational demands and accelerate model training, the cloud product grids in the training, validation, and test sets were cropped to 256 × 256 pixels, while the input TIR observations were resized to 128 × 128 pixels. Ultimately, the training set consisted of 76,247 samples, while the validation and test sets contained 9,530 and 5,740 samples, respectively.

To evaluate the performance of the cloud retrieval model, we employ cloud products from the active radar Cloud-Aerosol Lidar with Orthogonal Polarization (CALIOP) aboard the CALIPSO satellite. CALIPSO/CALIOP provides measurements at 0.532 $\mu$m and 1.064 $\mu$m, which are used to derive the vertical structure of clouds and aerosols, consistent with its designed mission \citep{winker2007}. In this study, we compare the model’s retrievals with CLP and CTH data obtained from the CALIPSO Level-2 cloud layer product. These data are derived from the feature classification flags and layer top altitude profiles, with a horizontal resolution of 1 km. Additionally, we further validate our results by comparing them with Himawari-8 operational cloud products (including CTH and CLP) released by the Japan Aerospace Exploration Agency (JAXA) P-Tree, to assess whether our retrieval methodology yields reasonable cloud properties.

\subsection{Diffusion model}
The diffusion model is a state-of-the-art deep learning technique that employs probabilistic denoising processes to develop generative models \citep{bishop2024}. The model typically operates on the principle of simulating a gradual process of denoising, effectively reconstructing data points from a noise-like distribution. This process is modeled as a reverse Markov chain, where a data sample is initially transformed into noise through a sequence of diffusion steps and then reconstructed back into a clean sample through learned reverse transitions. In a classical set-up, the model involves iteratively applying a series of conditional Gaussian distributions, beginning from a distribution of noise $ p(\mathbf{z}_T) $ and progressively denoising it to retrieve the original data distribution $ p(\mathbf{x}_0) $. This can be succinctly represented as,
\begin{equation}
p(\mathbf{x}_0) = \int \cdots \int p(\mathbf{x}_0 | \mathbf{x}_1) p(\mathbf{x}_1 | \mathbf{x}_2) \cdots p(\mathbf{x}_{T-1} | \mathbf{z}_T) p(\mathbf{z}_T) \, d\mathbf{x}_1 \cdots d\mathbf{x}_{T-1} d\mathbf{z}_T,
\end{equation}
Each reverse transition $p_(\mathbf{x}_{t-1}|\mathbf{x}_t)$ is modeled as a Gaussian distribution,
\begin{equation}
p(\mathbf{x}_{t-1} | \mathbf{x}_{t}) = \mathcal{N}\big(\mathbf{x}_{t-1}; \, \mu_{t-1}(\mathbf{x}_{t}), \, \sigma_{t-1}^2 \mathbf{I}\big).
\end{equation}
where $\mu_{t-1}(\mathbf{x}_{t})$ is the mean of the distribution, which is often predicted by a neural network based on $\mathbf{x}_{t}$ and the noise schedule; $\sigma_{t-1}^2$ is the variance for each time step, controlled by the variance schedule; and $\mathbf{I}$ is the identity matrix, representing isotropic Gaussian noise.

In each iteration, the model utilizes the noisy data from the previous step as input, subsequently refining it to a greater degree of accuracy in accordance with the data's original state. The denoising path is learned from training data, thereby enabling the model to effectively generate or reconstruct high-quality data samples.

\subsection{Conditional diffusion model}
\begin{figure}[!htbp]
\centering
\includegraphics[width=1\textwidth]{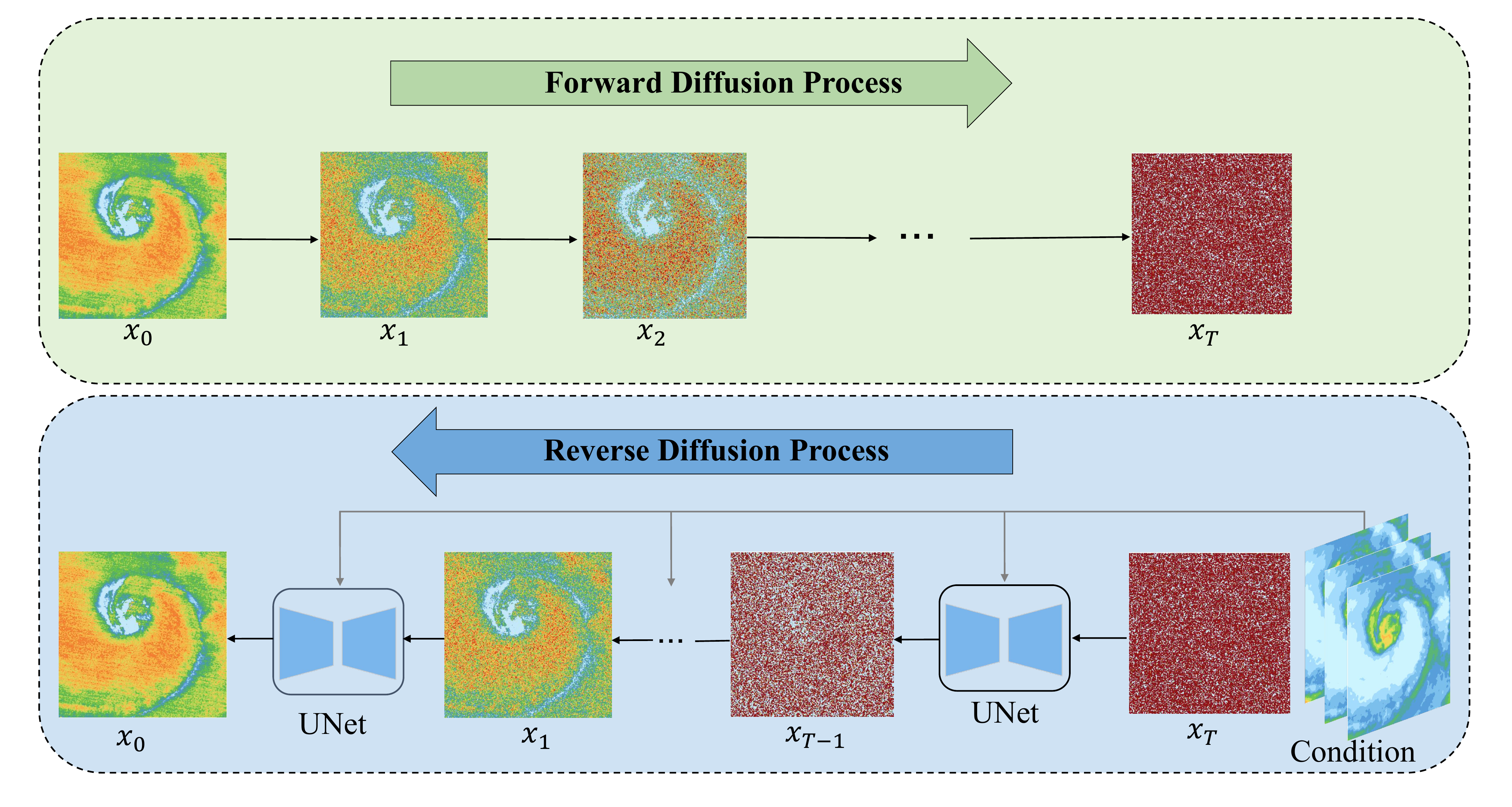}
\caption{The CloudDiff for high-resolution cloud identification and properties retrieval. The generated samples $\mathbf{x}$ are cloud  properties, and the conditions $\mathbf{y}$ includes TIR and VZA variables.}\label{fig6}
\end{figure}

In our study, these TIR measurements and VZA variable are denoted by $\mathbf{y}$, which is the condition variable. The target variables, cloud products, are represented by $\mathbf{x}$. The objective is to approximate the conditional distribution of $\mathbf{x}$ given $\mathbf{y}$, using a significantly large dataset of paired samples $(\mathbf{x}_i, \mathbf{y}_i)$. The conditional diffusion model incorporates conditioning variables into the generative process \citep{batzolis2021}, allowing the model to generate data conditioned on specific information. Mathematically, this can be represented as the transition from a noise distribution $ p(\mathbf{z}_T) $ to the data distribution $ p(\mathbf{x}_0) $ conditioned on a variable $ \mathbf{y} $, described by,
\begin{equation}
p(\mathbf{x}_0|\mathbf{y}) = \int p(\mathbf{x}_0|\mathbf{z}_T, \mathbf{y}) p(\mathbf{z}_T|\mathbf{y}) \, d\mathbf{z}_T,
\end{equation}
where, $ \mathbf{z}_T $ represents the latent variables at the final timestep, and the model iteratively refines these variables through the conditioning on $ \mathbf{y} $, enhancing its ability to target specific data generation tasks. As Figure~\ref{fig6} shows, the conditional diffusion model enables to produce cloud products given the conditions of TIR and VZA variables, making it particularly useful in scenarios where the output needs to be tailored to specific environments. In this framework, for any given $\mathbf{y}$, the algorithm outputs samples of $\mathbf{x}$ from $\mathbf{x} \sim p(\mathbf{x}_0|\mathbf{y})$, where $p$ is a learned distribution that does not adhere to any predefined probability distribution form. The forward process has the same scheme as the Denoising Diffusion Probabilistic Models (DDPMs) \citep{ho2020}, but in the reverse process we embed the conditional variables into the UNet for modelling the conditional probability distributions \citep{nai2024}.

In the forward process, the data $\mathbf{x}_0$ undergoes a series of transformations, gradually adding noise over discrete time steps $T$ until it is converted into pure Gaussian noise $\mathbf{x}_T \equiv \mathbf{z}_T$. The noise addition at each timestep $t$ is defined by a variance schedule $\beta_t$, and can be described by the following stochastic difference equation,
\begin{equation}
\mathbf{x}_t = \sqrt{1-\beta_t} \mathbf{x}_{t-1} + \sqrt{\beta_t} \mathbf{\epsilon}, \quad \mathbf{\epsilon} \sim \mathcal{N}(0, \mathbf{I}), 
\end{equation}
where $\mathbf{\epsilon}$ represents Gaussian noise.
By reparameterization~\cite{ho2020}, this process can also be written in closed form as,
\begin{equation}
\mathbf{x}_t = \sqrt{\bar{\alpha}_t} \, \mathbf{x}_0 + \sqrt{1-\bar{\alpha}_t} \, \mathbf{\epsilon}, 
\quad \bar{\alpha}_t = \prod_{s=1}^t (1-\beta_s),
\end{equation}

The reverse process, where the model learns to reconstruct the original data from noise, is explicitly conditioned on $\mathbf{y}$. 
At each step, the model learns the conditional probability distribution,
\begin{equation}
p_\theta(\mathbf{x}_{t-1}|\mathbf{x}_t, \mathbf{y}) 
= \mathcal{N}\!\Big(\mathbf{x}_{t-1}; \, \mu_\theta(\mathbf{x}_t, t, \mathbf{y}), \, \Sigma_\theta(\mathbf{x}_t, t)\Big),
\end{equation}
where the neural network parameterized by $\{ \theta \}$ predicts the mean $\mu_\theta(\mathbf{x}_t, t, \mathbf{y})$. Since $\mathbf{x}_t$ is available during training, the Gaussian noise term can be reparameterized so that the network predicts $\mathbf{\epsilon}$ instead of $\mu$, leading to
\begin{equation}
\mu_\theta(\mathbf{x}_t, t, \mathbf{y}) 
= \frac{1}{\sqrt{\alpha_t}} \Bigg(\mathbf{x}_t - \frac{1-\alpha_t}{\sqrt{1-\bar{\alpha}_t}} \, \mathbf{\epsilon}_\theta(\mathbf{x}_t, t, \mathbf{y}) \Bigg),
\end{equation}
Thus, the sampling step becomes
\begin{equation}
\mathbf{x}_{t-1} \sim \mathcal{N}\!\Bigg(\mathbf{x}_{t-1}; \, 
\frac{1}{\sqrt{\alpha_t}} \Big(\mathbf{x}_t - \frac{1-\alpha_t}{\sqrt{1-\bar{\alpha}_t}} \, \mathbf{\epsilon}_\theta(\mathbf{x}_t, t, \mathbf{y}) \Big), \, 
\Sigma_\theta(\mathbf{x}_t, t)\Bigg),
\end{equation}

The objective of training this conditional diffusion model is to minimize the difference between the estimated $\mathbf{x}_{t-1}$ and its actual value. This effectively allows the model to learn the reverse of the forward diffusion process. The loss function is originally from the Fisher divergence \citep{song2019,song2021,nai2024}, but equivalently used as a variant of the mean squared error between the predicted and actual previous timestep values, conditioned on $\mathbf{y}$, and this leads to a noise-prediction loss of the form,
\begin{equation}
  \mathcal{L}(\theta) = \mathbb{E}_{\mathbf{x}_0, \mathbf{\epsilon}, \mathbf{y}} \left[ \|\mathbf{\epsilon} - \mathbf{\epsilon}_{\theta}(\mathbf{x}_t, t, \mathbf{y})\|^2 \right],
\end{equation}
where $\mathbf{\epsilon}_{\theta}$ represents the outputs of the UNet as the predictions of the noise used to generate $\mathbf{x}_t$ from $\mathbf{x}_{t-1}$. To improve the representation ability, we have introduced the multi-head attention modules into the UNet architecture \citep{vaswani2017}. This formulation directly trains the network to predict the noise added at each timestep, thereby enabling effective denoising and sample generation. Note that we developed separate high-resolution models for each cloud property, because different cloud properties exhibit distinct characteristics. Moreover, training models separately allows each to serve as a general model for its respective cloud property. In future studies, each model can be fine-tuned based on more accurate cloud property products to fulfill required downstream tasks. 

After training, the conditional diffusion model (i.e., CloudDiff) is capable of generating multiple samples simultaneously. In our tests, we generated 30 samples per evaluation instance, taking into account time and computational resource constraints. These samples are reminiscent of the ensemble members used in dynamical models for numerical weather prediction, which employ large numbers of members for ensemble predictions \citep{li2024}. Furthermore, we conduct comparative analyses between CloudDiff and established deterministic data-driven methods. For this purpose, we use a supervised learning approach with a UNet architecture \citep{trebing2021}, referred to as the deterministic model, as the benchmark. This method is specifically applied to the tasks of high-resolution retrieval of cloud properties and CLP identification, serving as a baseline for performance comparison.

\subsection{Performance evaluation}
The CloudDiff serves as a retrieval approach that requires an appropriate evaluation scheme. Although sample-by-sample comparisons are intuitive, they cannot fully demonstrate the effectiveness of the retrieval technique. To obtain a comprehensive performance evaluation, we collect MODIS labels to assess the quality of the generated CLP and cloud properties. Consequently, we employ F1 score and OA to validate the retrieved CLP \citep{zhao2023}. Additionally, we employ MAE, RMSE, SSIM, and PSNR as metrics, allowing for a quantitative assessment of the model's performance in enhancing spatial resolution. These metrics, commonly used in cloud properties retrieval \citep{wang2022} and image similarity evaluation \citep{ling2024}—assess structural similarity and perceptual quality, are defined as follows,
\begin{equation}
\text{MAE} = \frac{1}{N N_p} \sum_{i=1}^N \sum_{j=1}^{N_p}\left|x_{i,j} - \hat{x}_{i,j} \right|,
\end{equation}
\begin{equation}
\text{RMSE} = \sqrt{\frac{1}{N N_p} \sum_{i=1}^N \sum_{j=1}^{N_p} (x_{i,j} - \hat{x}_{i,j})^2},
\end{equation}
 \begin{equation}
\text{SSIM} = \frac{(2\mu_x \mu_{\hat{x}} + C_1)(2\sigma_{x\hat{x}} + C_2)}{(\mu_x^2 + \mu_{\hat{x}}^2 + C_1)(\sigma_x^2 + \sigma_{\hat{x}}^2 + C_2)},
\end{equation}
\begin{equation}
\text{PSNR} = 10 \cdot \log_{10} \left( \frac{\text{MAX}^2}{\frac{1}{N N_p} \sum_{i=1}^N \sum_{j=1}^{N_p} (x_{i,j} - \hat{x}_{i,j})^2} \right),
\end{equation}
where $N$ represents the number of samples, $x_i$ denotes the values from MODIS/CALIPSO cloud products, and $ \hat{x}_i $ represents the high-resolution retrieved cloud products. $N_p$ indicates the number of pixels for each sample, and $j$ labels the index of the pixels.  $\mu_x$ and $\mu_{\hat{x}}$ denote the mean values of $x$ and $\hat{x}$, respectively; $\sigma_x^2$ and $\sigma_{\hat{x}}^2$ denote their variances; $\sigma_{x\hat{x}}$ denotes the covariance between $x$ and $\hat{x}$; $C_1$ and $C_2$ are small constants to stabilize the division. $\text{MAX}$ is the maximum possible pixel value of the data. It should be noted that a more accurate retrieval model will have smaller RMSE and MAE, larger SSIM, and higher PSNR.

\backmatter

\section*{Data availability}
The authors would like to thank JAXA for freely providing the Himiwari-8/AHI observations and cloud products (\url{https://www.eorc.jaxa.jp/ptree/}), NASA for freely providing the MODIS cloud products and precipitation data online (\url{https://ladsweb.modaps.eosdis.nasa.gov/}), National Meteorological Center of the China Meteorological Administration for freely providing best track of typhoon (\url{http://typhoon.nmc.cn/}). We acknowledge Xiaoye Wang from Fudan University for assisting with data processing.

\section*{Acknowledgements}
This work was supported by the National Natural Science Foundation of China (42222506 and 42450254). L. Wang thanks the National Natural Science Foundation of China (12147101) for supporting his visit to Fudan University. L. Wang is also supported by JSPS KAKENHI Grant No. 25H01560 and JST-BOOST Grant No.JPMJBY24H9. We thank the DEEP-IN working group at RIKEN-iTHEMS for support in the preparation of this study. 

\section*{Author Contributions}
H.X., and F.Z. designed the project.  H.X., F.Z., L.W., and B.P., designed and performed the model evaluation. H. X. performed the analysis under supervision of F.Z., L.W., B.P., Y.Z., M.W., and J.L.. H.X., L.W., W.L., and B.G. wrote and revised the manuscript. F.Z., L.W., B.P.,Y.Z., M.W., and J.L. contributed to interpreting results and discussions.
 
\section*{Declarations}
The authors declare that they have no known competing financial interests or personal relationships that could have appeared to influence the work reported in this paper.


\bibliography{reference}


\section*{Supplementary Materials}
\vspace{1cm}
This Supplementary Material file includes: \\
Table S1 \\
Figs. S1 to S5 

\vspace{1cm}
\renewcommand{\thefigure}{S\arabic{figure}}
\renewcommand{\thetable}{S\arabic{table}}
\setcounter{figure}{0} 

\begin{table}[!htbp]
\centering
\caption{Evaluating CloudDiff’s performance: a comparison of ensemble mean, ensemble median, and maximum a posteriori estimation based on kernel density estimation.}
\label{tabS1}
\begin{tabular}{lcccccccccc}
\toprule
\multirow{2}{*}{Category} & \multirow{2}{*}{Evaluation metrics} & \multicolumn{3}{c}{COT} & \multicolumn{3}{c}{CER ($\mu$m)} & \multicolumn{3}{c}{CTH (km)} \\
\cmidrule(lr){3-5} \cmidrule(lr){6-8} \cmidrule(lr){9-11}
 & & Mean & Median & MAP & Mean & Median & MAP & Mean & Median & MAP \\
\midrule
\multirow{2}{*}{ALL} & MAE & 6.12 & 6.19 & 6.43 & 5.71 & 5.96 & 6.30 & 1.30 & 1.31 & 1.37 \\
 & RMSE & 12.43 & 13.14 & 13.89 & 8.83 & 9.89 & 10.53 & 2.32 & 2.65 & 2.80 \\
\addlinespace
\multirow{2}{*}{Liquid} & MAE & 6.67 & 6.92 & 7.26 & 4.12 & 4.38 & 4.66 & 1.06 & 1.08 & 1.12 \\
 & RMSE & 12.73 & 13.50 & 14.20 & 6.18 & 7.11 & 7.60 & 1.95 & 2.19 & 2.33 \\
\addlinespace
\multirow{2}{*}{Ice} & MAE & 5.58 & 5.45 & 5.60 & 7.31 & 7.54 & 7.94 & 1.60 & 1.62 & 1.69 \\
 & RMSE & 12.13 & 12.76 & 13.56 & 10.87 & 12.05 & 12.82 & 2.75 & 3.15 & 3.32 \\
\bottomrule
\end{tabular}
\end{table}

\begin{figure}[!htbp]
\centering
\includegraphics[width=1\linewidth]{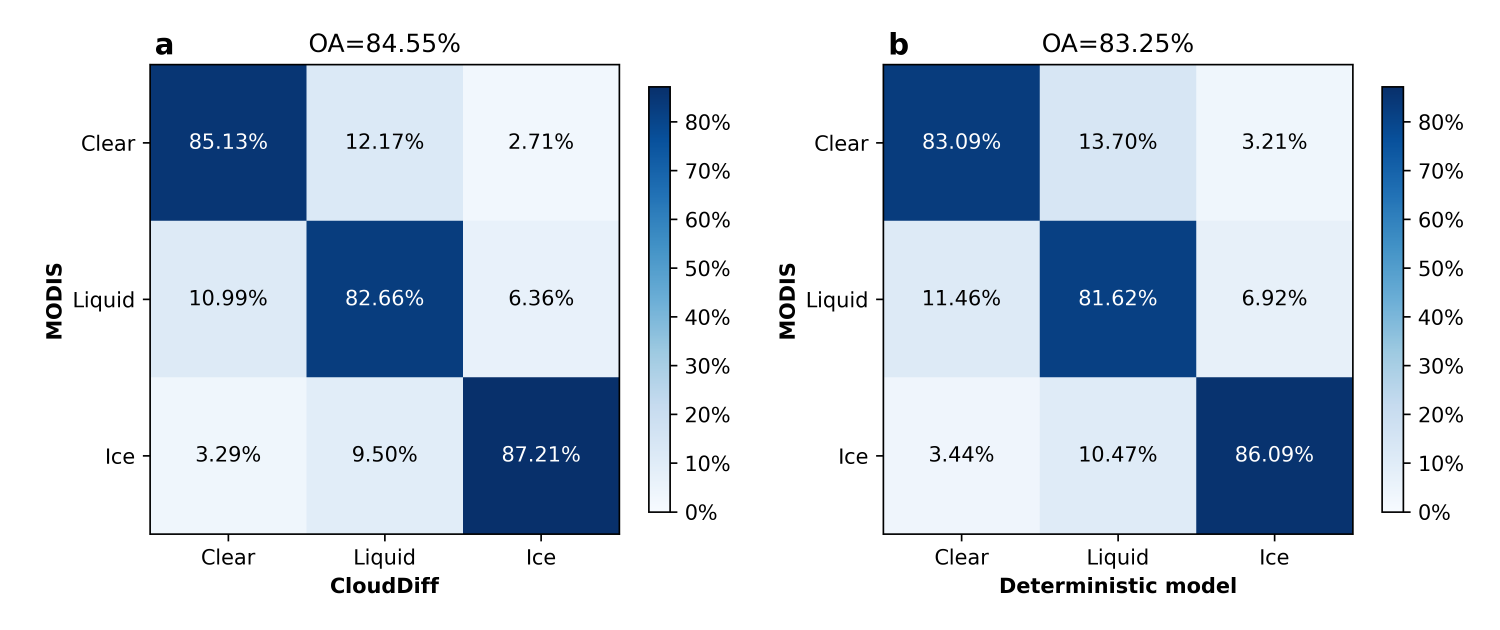}
\caption{Confusion matrices of CLP products comparing MODIS with (a) CloudDiff (ensemble mean of 30 sample sizes) and (b) the deterministic model.}
\label{figS1}
\end{figure}

\vspace{1cm}

\begin{figure}[!htbp]
\centering
\includegraphics[width=1\linewidth]{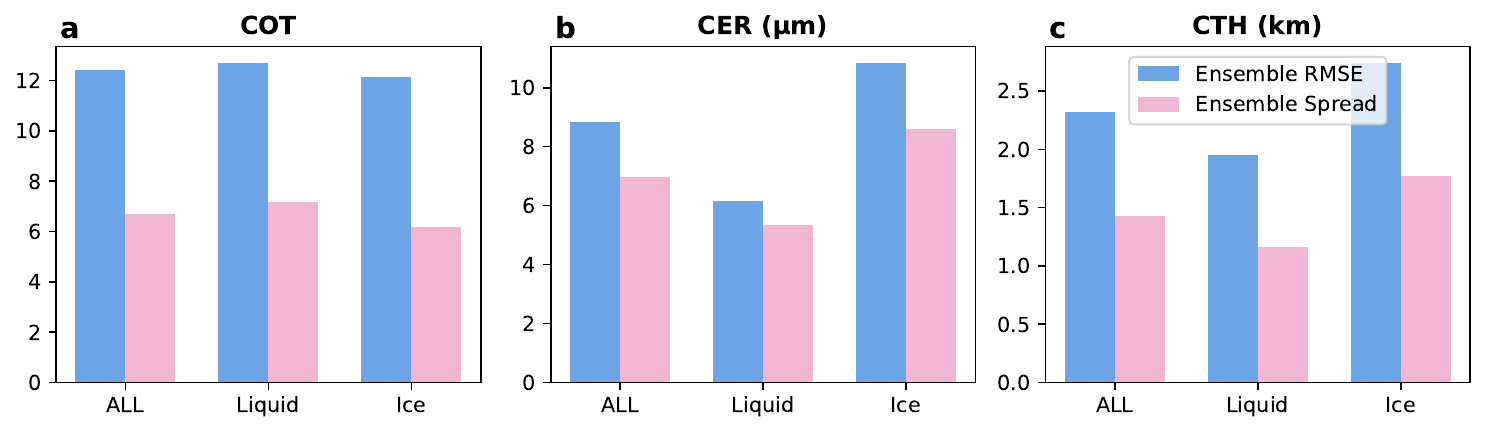}
\caption{Ensemble RMSE and ensemble spread of retrieved cloud properties by CloudDiff. Skill metrics were calculated between the CloudDiff model and MODIS cloud products.}
\label{figS2}
\end{figure}

\vspace{1cm}

\begin{figure}[!htbp]
\centering
\includegraphics[width=1\linewidth]{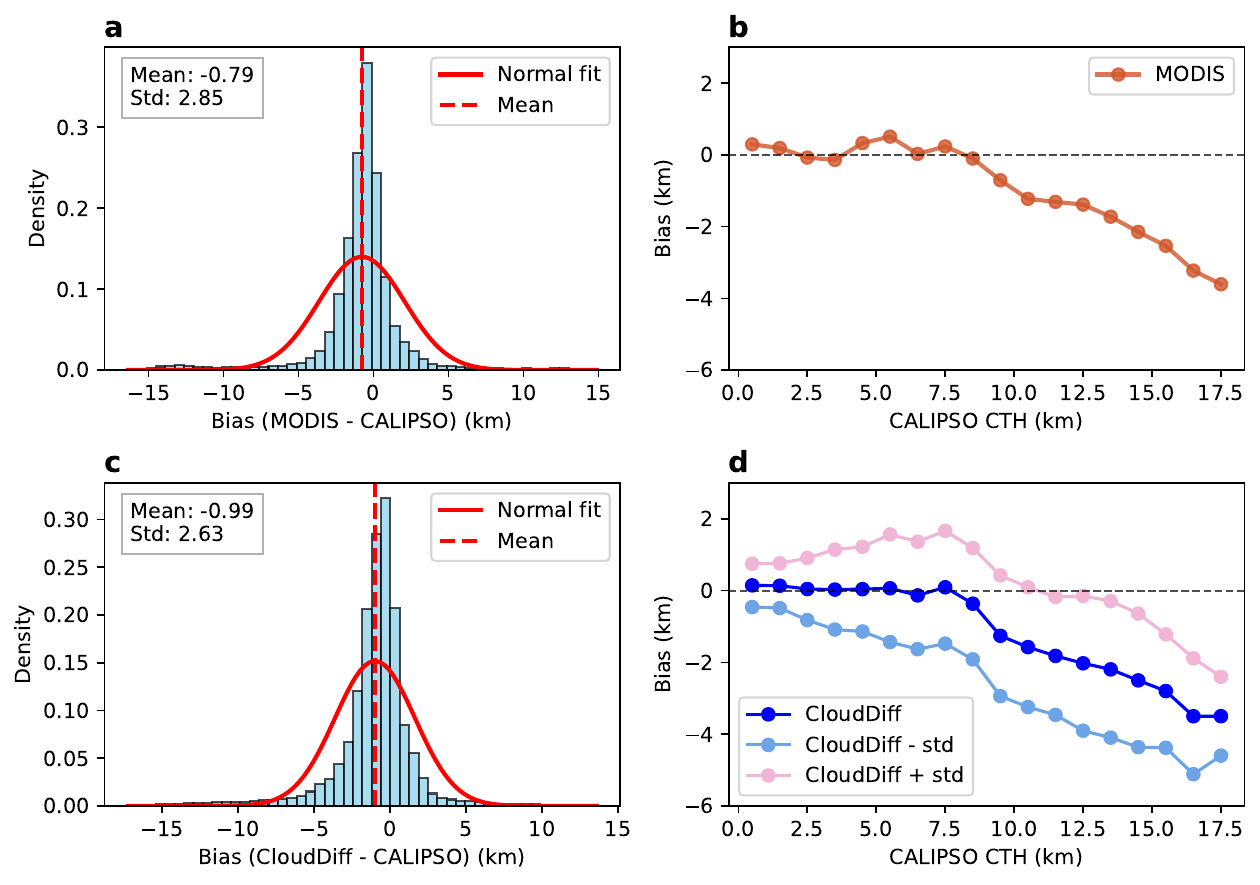}
\caption{Performance evaluation of cloud properties retrieved by the CloudDiff and deterministic model across different satellite viewing zenith angles (VZA). (a–d) show the results for COT, CER, CTH, and CLP, respectively.}
\label{figS3}
\end{figure}

\vspace{1cm}

\begin{figure}[!htbp]
\centering
\includegraphics[width=1\linewidth]{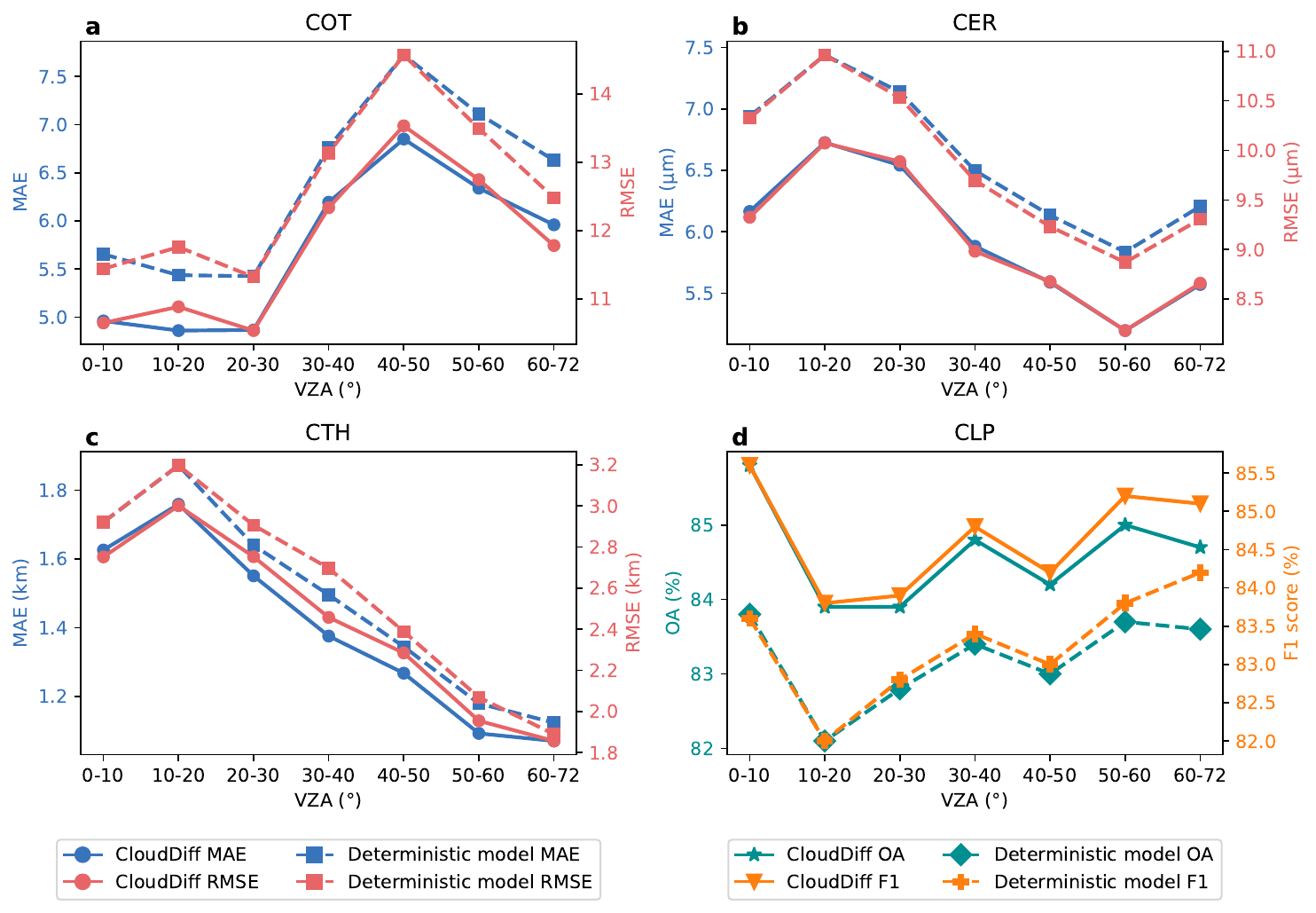}
\caption{Comparison of MODIS CTH and CloudDiff CTH biases relative to CALIPSO CTH, where bias is defined as retrieval minus CALIPSO. (a) and (c) present histograms of the biases with its mean and standard deviation (Std), along with fitted normal distributions. (b) and (d) show the variation of bias as a function of CALIPSO CTH for MODIS and CloudDiff retrievals, including CloudDiff bias estimates shifted by ±1 Std.}
\label{figS4}
\end{figure}

\vspace{1cm}

\begin{figure}[!htbp]
\centering
\includegraphics[width=1\linewidth]{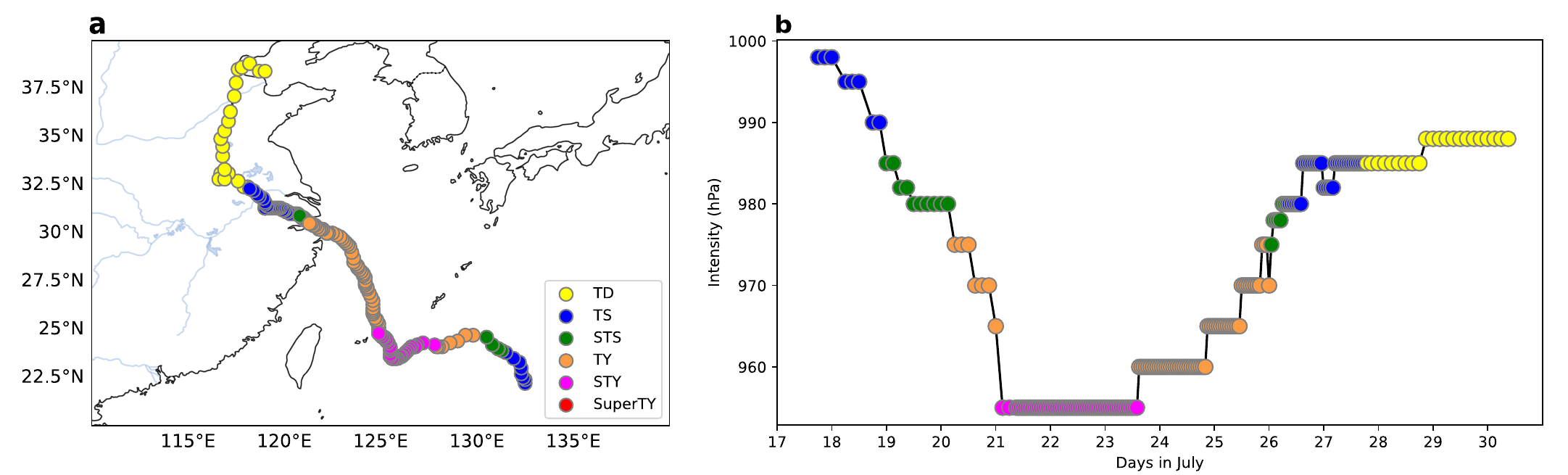}
\caption{The best track of typhoon In-Fa (No. 2106) (a), and the time series of the minimum central sea level pressure at the center of typhoon In-Fa (b). Note that TD, TS, STS, TY, STY, and SuperTY represent tropical depression, tropical storm, severe tropical storm, typhoon, severe typhoon, and super typhoon, respectively.}
\label{figS5}
\end{figure}

\restoregeometry
\end{document}